\documentclass[prb,twocolumn]{revtex4}

\usepackage{graphicx}
\usepackage{dcolumn}
\usepackage{bm}
\newcommand{\be}{\begin{equation}}
\newcommand{\ee}{\end{equation}}

\begin{document}

\draft
\title
{\bf  Tunneling current spectroscopy of a nanostructure junction
involving multiple energy levels}

\author{David M.-T. Kuo$^1$,Yia-Chung Chang$^{2,3}$}
\email{ychang15@uiuc.edu} \affiliation{{}$^1$Department of
Electrical Engineering, National
Central University, Chung-Li, Taiwan 320\\
$^2$Research Center for Applied Sciences, Academia Sinica, Taipei,
Taiwan 115\\
$^3$Department of Physics, University of Illinois at
Urbana-Champaign, Urbana, Illinois 6180}

\date{\today}

\begin{abstract}
A multi-level Anderson model is employed to simulate the system of a
nanostructure tunnel junction with any number of one-particle energy
levels. The tunneling current, including both shell-tunneling and
shell-filling cases, is theoretically investigated via the
nonequilibrium Green's function method. We obtain a closed form for
the spectral function, which is used to analyze the complicated
tunneling current spectra of a quantum dot or molecule embedded in a
double-barrier junction. We also show that negative differential
conductance can be observed in a quantum dot tunnel junction when
the Coulomb interactions with neighboring quantum dots are taken
into account.
\end{abstract}

\maketitle

In the last decade, the tunneling current spectra of nanostructure
junctions such as single-electron transistors (SETs)$^{1-2}$ and
molecule transistors (MTs)$^{3-5}$ have been extensively studied due
to their important applications in quantum computing, quantum
communication, and ultrahigh-density IC circuits. Although the
single-particle energy levels of such systems can be obtained by
{\it ab inito} or semiempirical methods$^{6-10}$, it is still
difficult to model the full tunneling current spectra due to the
presence of Coulomb blockade and the uncertainty in dot shape, size
and position.

The characteristics of tunneling current of SETs or MTs can be
classified into either the shell-tunneling case (with no charge
accumulation) or the shell-filling case (with charge accumulation).
In the shell-tunneling case, the electron Coulomb interactions are
suppressed. Consequently, current spectra directly reveals the
one-particle resonance energy levels. In the shell-filling case, the
tunneling current spectra become much more complicated due to the
presence of electron Coulomb interactions. Consequently, the
charging energies and energy levels can not be readily and directly
obtained from experimental data. Both physical parameters are
crucial not only in the optimization of tunneling devices, but also
for the understanding of fundamental physics of miniaturized
nanostructures. Thus, it is desirable to derive a simple formula
that allows one to determine the multiple energy levels involved and
the associated Coulomb interactions from the measured tunneling
current spectra.

In this letter, we derive a simple yet general analytic expression
that is applicable for modeling the tunneling current through any
nanostructure (including QDs and molecules) involving multiple
energy levels (i.e. the ground state plus all the excited states of
interest). This is done by solving a multi-level Anderson model via
non-equilibrium Green's function technique,$^{11}$ which has been
extensively used for investigating the Coulomb blockade and Kondo
effect on the tunneling current through the ground state of a single
QD.$^{12}$ We find that when more than one level in a QD/molecule
are involved in the charge transport, the average two-particle
occupation numbers play an important role in determining whether a
given energy channel is close or open to the tunneling current.

The system of an isolated nanostructure embedded in a double-barrier
junction can be described by the following Hamiltonian $^{13-16}$:
\begin{eqnarray}
H& =&\sum_{k,\sigma,\beta} \epsilon_k
a^{\dagger}_{k,\sigma,\beta}a_{k,\sigma,\beta}+\sum_{\ell,\sigma}
E_{\ell} d^{\dagger}_{\ell,\sigma} d_{\ell,\sigma}\nonumber \\
\nonumber &+&\sum_{\ell,j,\sigma,\sigma'} U_{\ell,j}
d^{\dagger}_{\ell,\sigma} d_{\ell,\sigma} d^{\dagger}_{j,\sigma'}
d_{j,\sigma'}+\sum_{k,\sigma,\beta,\ell}
V_{k,\beta,\ell}a^{\dagger}_{k,\sigma,\beta}d_{\ell,\sigma}\\
 &+&\sum_{k,\sigma,\beta,\ell}
V^{*}_{k,\beta,\ell}d^{\dagger}_{\ell,\sigma}a_{k,\sigma,\beta},
\end{eqnarray}
where $a^{\dagger}_{k,\sigma,\beta}$ ($a_{k,\sigma,\beta}$) creates
(destroys) an electron of momentum $k$ and spin $\sigma$ with energy
$\epsilon_k$ in the $\beta$ metallic electrode.
$d^{\dagger}_{\ell,\sigma}$ ($d_{\ell,\sigma}$) creates (destroys)
an electron inside the nanostructure with orbital energy $E_{\ell}$.
$U_{\ell,j}$ describes the intra- or inter-level charging energies,
and $V_{k,\beta,\ell}$ describes the coupling between the band
states in the contacts and the confined states in the nanostructure.

Using Keldysh Green's function technique$^{11,12}$, we obtain the
tunneling current through a nanostructure,
\begin{equation}
J=\frac{-2e}{\hbar}\sum_{\ell} \int \frac{d\epsilon}{2\pi}
\frac{(f_{L}-f_R)\Gamma_{\ell,L}(\epsilon)
\Gamma_{\ell,R}(\epsilon)}
{\Gamma_{\ell,L}+\Gamma_{\ell,R}}ImG^r_{\ell,\sigma},
\end{equation}
where $f_{L}= f(\epsilon-\mu_{L})$ and $f_{R}=f(\epsilon-\mu_{R})$
are the Fermi distribution functions for the left and right
electrode, respectively. $G^r_{\ell,\sigma}(\epsilon)$ is the
retarded Green's function. The chemical potential difference between
these two electrodes is equal to the applied bias $eV_a$.
$\Gamma_{\ell,L}$ and $\Gamma_{\ell,R}$
[$\Gamma_{\ell,\beta}=\sum_{{\bf k}} |V_{\ell,\beta,{\bf k}}|^2
\delta(\epsilon-\epsilon_{{\bf k}})]$ denote the tunneling rates
from the nanostructure to the left (source) and right (drain)
electrodes, respectively. For simplicity, these tunneling rates will
be assumed to be energy independent. Therefore, the calculation of
tunneling current is entirely determined by the spectral function,
$A(\epsilon)=ImG^r_{\ell,\sigma}(\epsilon)$.

Rigorous solution to $G^{r}_{\ell,\sigma}$ in the Coulomb blockade
regime can be obtained by solving a hierarchy of equations of
motion, which relate $G^{r}_{\ell,\sigma}$ to two-particle Green's
functions, which can be solved due to the termination of the series
at some finite order. After tedious algebra, we obtain a neat closed
form (by keeping the two-particle correlation functions associate
with the same level)
\begin{eqnarray}
G^{r}_{\ell,\sigma}(\epsilon)&=&(1-N_{\ell.-\sigma})\sum^{3^{n-1}}_{m=1}
\frac{p_m}{\epsilon-E_{\ell}-\Pi_m+i\frac{\Gamma_{\ell}}{2}} \nonumber \\
 &+& N_{\ell.-\sigma}\sum^{3^{n-1}}_{m=1}
\frac{p_m}{\epsilon-E_{\ell}-U_{\ell}-\Pi_m+i\frac{\Gamma_{\ell}}{2}},
\end{eqnarray}
where $n$ denotes the number of one-particle energy levels
considered for the nanostructure. $\Pi_m$ denotes the sum of Coulomb
interactions seen by a particle in level $\ell$ due to other
particles in configuration $m$, in which each level $j (j\ne \ell)$
can be occupied by zero, one or two particles. $p_m$ denotes the
probability of finding the system in configuration $m$. For a
two-level ($n=2$) system ($\ell \neq j$), we have three
configurations with $p_1=a^j\equiv 1-(N_{j,\sigma}+N_{j,-\sigma})+
\langle n_{j,\sigma}n_{j,-\sigma}\rangle$ (the probability with no
particle in level $j$), $p_2=b^j\equiv N_{j,\sigma}+N_{j,-\sigma}-2
\langle n_{j,\sigma}n_{j,-\sigma}\rangle$ (the probability with one
particle in level $j$), and $p_3=c^{j} \equiv \langle
n_{j,\sigma}n_{j,-\sigma}\rangle$ (the probability with two
particles in level $j$). Meanwhile, $\Pi_1=0$, $\Pi_2=U_{\ell j}$
and $\Pi_3=2U_{\ell j}$. $U_{\ell j}$ denotes the interlevel Coulomb
interaction between an electron at level $\ell$ and the other at
level $j$. For a three-level system ($\ell \neq j \neq j'$), there
are nine ($3\times 3$) configurations, and the probability factors
become $p_1=a^j a^{j'}$, $p_2=b^j a^{j'}$, $p_3=a^{j} b^{j'}$,
$p_4=c^j a^{j'}$, $p_5=c^{j'}a^{j}$, $p_6=b^j b^{j'}$, $p_7=c^j
b^{j'}$, $p_8=c^{j'} b^{j}$, and $p_9=c^{j} c^{j'}$. Interlevel
Coulomb interaction factors are $\Pi_1=0$, $\Pi_2=U_{\ell j}$,
$\Pi_3=U_{\ell j'}$, $\Pi_4=2U_{\ell j}$, $\Pi_5=2U_{\ell j'}$,
$\Pi_6=U_{\ell j}+U_{\ell j'}$, $\Pi_7=2U_{\ell j}+U_{\ell j'}$,
$\Pi_8=2U_{\ell j'}+U_{\ell j}$, and $\Pi_9=2U_{\ell j}+2U_{\ell
j'}$. Based on these simple rules , the probability factors and
interlevel Coulomb interaction factors for any number of energy
levels can be similarly determined.
We see that the probability of
finding the system in each configuration is determined not only by
the average one-particle occupation numbers but also by the average
two-particle occupation numbers. $N_{\ell,\sigma}$ in Eq. (3) can be
obtained by solving the following equations self-consistently
\begin{small}
\begin{equation}
N_{\ell,\sigma} = -\int \frac{d\epsilon}{\pi} \frac{\Gamma_{\ell,L}
f_{L}+\Gamma_{\ell,R} f_{R}}{\Gamma_{\ell,L}+\Gamma_{\ell,R}}
ImG^r_{\ell,\sigma}(\epsilon),
\end{equation}
\end{small}
\begin{eqnarray}
\langle n_{\ell,\sigma}n_{\ell,-\sigma}\rangle = -\int
\frac{d\epsilon}{\pi} \frac{\Gamma_{\ell,L} f_{L}+\Gamma_{\ell,R}
f_{R}}{\Gamma_{\ell,L}+\Gamma_{\ell,R}} Im
G^{r}_{\ell,\ell}(\epsilon),
\end{eqnarray}
\begin{eqnarray}
G^{r}_{\ell,\ell}(\epsilon)=N_{\ell.-\sigma}\sum^{3^{n-1}}_{m=1}
\frac{p_m}{\epsilon-E_{\ell}-U_{\ell}-\Pi_m+i\frac{\Gamma_{\ell}}{2}}.
\end{eqnarray}
Both $N_{\ell,\sigma}$ and $N_{\ell,\ell}\equiv \langle
n_{\ell,\sigma}n_{\ell,-\sigma}\rangle$ are limited to the range
between 0 and 1.

To illustrate the usefulness of our theory, we calculate the
tunneling current spectra of an isolated CdSe QD sandwiched between
an STM tip (left lead) and a conducting substrate (right lead) as
studied experimentally in Ref.~\onlinecite{Jdi}. We consider the
case of three energy levels. Because the tip-substrate is biased at
$V_a$, the bare energy levels of $E_{\ell}$ in the dot are changed
to $E_{\ell} + \alpha eV_a$ . The factor
$\alpha=L_{L}/(L_{L}+L_{R})$ is determined by the distances from the
QD center to the tip ($L_L$) and the substrate ($L_R$),
respectively. Here $\alpha=0.61$, which is determined from the
separation between the first peak at negative bias and that at
positive bias.\cite{Jdi} Other physical parameters can also be
determined by comparing Eq. (3) with the peak positions of the
tunneling spectra observed in Ref.~\onlinecite{Jdi}. We obtain the
following. The chemical potentials of both electrodes are $0.78$~eV
below the ground state level $E_1$ at zero bias. $E_2-E_1=0.236$~eV,
$E_3-E_1=0.456$~eV, $U_{1}=0.137$~eV, $U_{12}=U_{21}=0.122$~eV,
$U_{2}=0.07$~eV, $U_3=0.06$~eV, $U_{13}=U_{31}=0.1$~eV, and
$U_{23}=U_{32}=0.04$~eV. These intra- and inter-level Coulomb
energies are reasonable for the $~3$ nm diameter CdSe QD considered
here.  We note that the p-like level in a spherical QD is sixfold
degenerate (including spin). However, the coupling strength between
the tip (or substrate) and the $p_x$- or $p_y$-like orbital is weak.
Therefore, only $p_z$-like orbital has been included, which is
labeled by $E_2$, while $E_3$ denotes one of the d-like orbitals
that is strongly coupled to the leads.

Occupation numbers $N_{\ell}$ and $N_{\ell,\ell}$ are obtained by
solving the coupled equations Eqs.~(4)-(6). Once they are
determined, we can calculate the tunneling current by substituting
Eq. (3) into Eq. (2). Figure 1 shows the differential conductance,
$dJ/dV_a$ as a function of applied bias for various ratios of
$\Gamma_{L}/\Gamma_{R}$. Curve (c) exhibits the maximum number of
resonant channels, which correspond to the poles of the retarded
Green's function in Eq. (3). The first eight peaks of curve(c) (with
energy less than $E_3$) are $\epsilon_1=E_1$, $\epsilon_2=E_1+U_1$,
$\epsilon_3=E_2$,~$\epsilon_4=E_1+U_1+U_{12}$,~$\epsilon_5=E_2+U_2$,
$\epsilon_6=E_2+U_{12}$,~$\epsilon_7=E_1+U_1+2U_{12}$, and
$\epsilon_8=E_2+U_2+U_{12}$. The charging energy of the ground
state, $U_1$ can be determined by the difference in bias between the
first two peaks. The two channels, $\epsilon=E_1+U_{12}$ and
$\epsilon=E_1+2U_{12}$ are prohibited because those resonance
energies are below $E_2$ (hence $N_2=0$ and $N_{22}=0$ when the
applied bias aligns with these channels). The limit
$\Gamma_{L}/\Gamma_{R} \ll 1$ corresponds to the so called
"shell-tunneling case".$^{17}$ According to Eqs.~(4) and (5),
$N_{\ell}$ and $N_{\ell,\ell}$ are determined by the factor ${\cal
P}=\Gamma_{\ell,L}/(\Gamma_{\ell,L}+\Gamma_{\ell,R})$. Therefore,
$N_{\ell} $ and $N_{\ell,\ell}$ approaches zero in this limit.
Consequently, the peaks associated with the resonance levels
involving the intralevel and interlevel Coulomb interactions will be
suppressed, since the carrier will tunnel out much faster than it
can be fed. The peaks associated with the single-particle levels,
$E_1$, $E_2$ and $E_3$ will be the only resonance channels allowed
in the limit of $\Gamma_{L}/\Gamma_{R} \ll 1$ as seen in curve (e).
In the opposite limit, $\Gamma_{L}/\Gamma_{R} \gg 1$ we have the so
called "shell-filling case" [see curve(a)]. In this case, both $N_1$
and $N_{11}$ approach 1  when $V_a> E_1 +U_1$.  Consequently,
resonance channels related at $E_2$ and $E_2+U_2$ are suppressed
while the resonances at $E_2+2U_{12}$ and $E_2+U_2+2U_{12}$ are
enhanced. This indicates that the bare energy levels are
renormalized if charges reside in the QD. Similarly for $E_3$
related channels.

Next, we compare our theoretical prediction for the tunneling
current spectra with those measured in Ref.~\onlinecite{Jdi}, where
the observed differential conductance peaks are broadened. The
broadening is mainly due to coupling to nearby QDs since the
experiment was performed at $T=4 K$. To take into account the above
broadening effect, we replace each Lorentzian function appearing in
the differential conductance by a Gaussian function of the form $f_i
\exp\{-\frac{(\epsilon-\epsilon_i)^2}{2\rho_{i}^2}\}/(\rho_i
\sqrt{2\pi})$. $f_i$, $\epsilon_i$ and $\rho_i$ denote the peak
strength, resonance energy,  and broadening width, respectively.
Figure 2 shows the predicted differential conductance as a function
of applied bias at zero temperature.  We have used the following
parameters $\Gamma_{L,1}=1$~meV, $\Gamma_{R,1}=0.15$~meV,
$\Gamma_{L,2}=3$~meV, $\Gamma_{R,2}=0.6$~meV.$
\Gamma_{L,3}=1.5$~meV, $\Gamma_{R,3}=0.375$~meV. The variation in
tunneling rates for different levels reflects the difference in wave
functions and the bias-dependent barrier height. For simplicity, we
assume $\rho_i=\rho+\Gamma_i$, where $\rho=35$~meV characterizes the
broadening due to coupling to nearby QDs. Both the positions and
relative strengths of these peaks are in very good agreement with
the experimental measurement reported in Ref.~\onlinecite{Jdi}. If
we remove the extrinsic broadening by setting $\rho=0$, the
differential conductance spectrum exhibits more resonance channels
as shown in the lower part of the figure, which can not be resolved
experimentally.

To illustrate the significance of the two-particle occupation
number, we apply our theory to the case of a coupled dot embedded
between two leads, where two QDs (dot A and dot B) are coupled by
interdot Coulomb interactions, even though the inter-dot tunneling
is weak. It is assumed that the right lead is closer to dot A than
to dot B. Thus, the tunneling rate $\Gamma_{R,B}$ is ten times
smaller than $\Gamma_{R,A}$, and the total current is dominated by
the current through dot A ($J_A$).
Figure 3 shows the total tunneling current, $J=J_A+J_B$ for various
strengths of the interdot Coulomb interaction (which may, for
example, correspond to different separations between the two dots).
It is seen that negative differential conductance (NDC) can occur
due to the interdot Coulomb interaction. The NDC characteristic of
tunneling current can be understood mainly from the feature of
$J_A$, since $J_A \gg J_B$.  NDC first occurs at $E_B$ as dot B
becomes filled, which leads to a suppression of the current through
the channel $E_A$, and the current resumes as the channel
$E_A+U_{AB}$ opens up when the bias further increases. For
comparison, we also plot in Fig.~3 the total current, $J_a+J_b$
obtained with the approximation,
$<n_{j,\sigma}n_{j,-\sigma}>=N_{j,\sigma} N_{j,-\sigma}$
(dash-dotted curve). The difference between this curve and the solid
curve shows the significance of treating the two-particle occupation
number correctly. In particular, we see in the dash-dotted curve
($J_a+J_b$) that the resonance channel at $\epsilon=E_A+2U_{AB}$ is
allowed even though the applied bias is below $E_B+U_B$ (i.e. dot B
remains singly charged). Such a nonphysical behavior in $J_a+J_b$
arises due to the approximation $N_{BB}\approx N_B N_B \ne 0$ used,
while if treated correctly $N_{BB}$  should vanish under this
condition.
We also notice that the tunneling current increases only slightly
when applied bias overcomes the charging energy of dot A (marked by
$E_A+U_A$) since we are considering a shell-filling case for dot B
($\Gamma_{L,B}/\Gamma_{R,B}=10$) here ($N_{B} \approx 1$).
Increasing the applied bias beyond the resonance channel at
$E_A+U_A+U_{AB}$ leads to a significant jump in the tunneling
current. However, this tunneling current is suppressed again at
$E_B+U_B+U_{AB}$ because dot B is now filled with two charges.

To further analyze the NDC behavior due to interdot Coulomb
interaction, we plot in Figure 4 the differential conductances for
various ratios of $\Gamma_{L,B}/\Gamma_{R,B}$. Curve (a) corresponds
to the derivative of the curve shown in Fig.~3. The NDC behavior is
quite apparent. However, the NDC behavior disappears in curves (b)
and (c) when dot B is not under the shell-filling condition, even
though the interdot Coulomb energy remains the same. When dot B is
in the shell tunneling limit ($\Gamma_{L,B} \ll \Gamma_{R,B}$),
$J_A$ becomes unaffected by dot B, and the tunneling current [curve
(c)] exhibits mainly two resonance channels at $E_A$ and $E_A+U_A$
plus a weaker peak at $E_B+U_{AB}$, which is contributed from $J_B$.
As illustrated in Figs.~3 and 4, when one employs an SET as a charge
detector, the interdot Coulomb interactions due to neighboring QDs
and the tunneling rate ratios, $\Gamma_{L}/\Gamma_{R}$ should be
carefully considered in the design.

   In conclusion, a closed form expression for
the spectral function for nanostructure tunnel junctions involving
multiple energy levels has been derived and it has be employed to
analyze the complicated spectra of the tunneling current through a
realistic single QD.  Our present theory also provides a useful
guideline for the design of SET with coupled quantum dots. We show
that when the interdot hoping effect is weak but the interdot
Coulomb interactions are strong, NDC behavior can occur and it
depends sensitively on the tunneling rate ratios for incoming and
outgoing electrons in each QD.


This work was supported by the National Science Council of the
Republic of China under Contract No. NSC-95-2221-E-008-147 and by
the Academia Sinica.


Figure captions

Fig. 1 Differential conductance as a function of applied bias at
for various ratios of $\Gamma_{L}/\Gamma_{R}$: (a) $10$, (b) $5$,
(c) $1$, (d) $0.4$, and (e) $0.2$. $\Gamma_{R}$ is fixed at 1~meV.

Fig. 2 Differential conductance as a function of applied bias with
and without the extrinsic broadening effect due to coupling to
nearby QDs. For better display, the broadened spectra has been
multiplied by a factor 5.

Fig. 3 Tunneling current ($J_A+J_B$) as a function of bias for
various strengths of interdot Coulomb interaction, $U_{AB}$. The
parameters used are: $E_A=E_F+108~$meV, $E_B=E_A+10~$meV,
$U_A=111~meV$, $U_B=108$~meV,
$\Gamma_{L,A}=\Gamma_{R,A}=\Gamma_{L,B}=1$~meV,
 and $\Gamma_{R,B}=0.1$~meV, and $\alpha=0.61$. Current is in
units of $J_0$=e*(meV)/h. Dot-dashed curve is for $J_a+J_b$ with
$U_{AB}=30~$meV.

Fig. 4 Differential conductance as a function of applied bias for
various tunneling rate ratios of $\Gamma_{L,B}/\Gamma_{R,B}$.
$U_{AB}=30$~meV. (a) $\Gamma_{L,B}=1$~meV and
$\Gamma_{R,B}=0.1$~meV.  (b) $\Gamma_{L,B}=1$~meV and
$\Gamma_{R,B}=1$~meV. (c) $\Gamma_{L,B}=0.1$~meV and
$\Gamma_{R,B}=1$~meV.

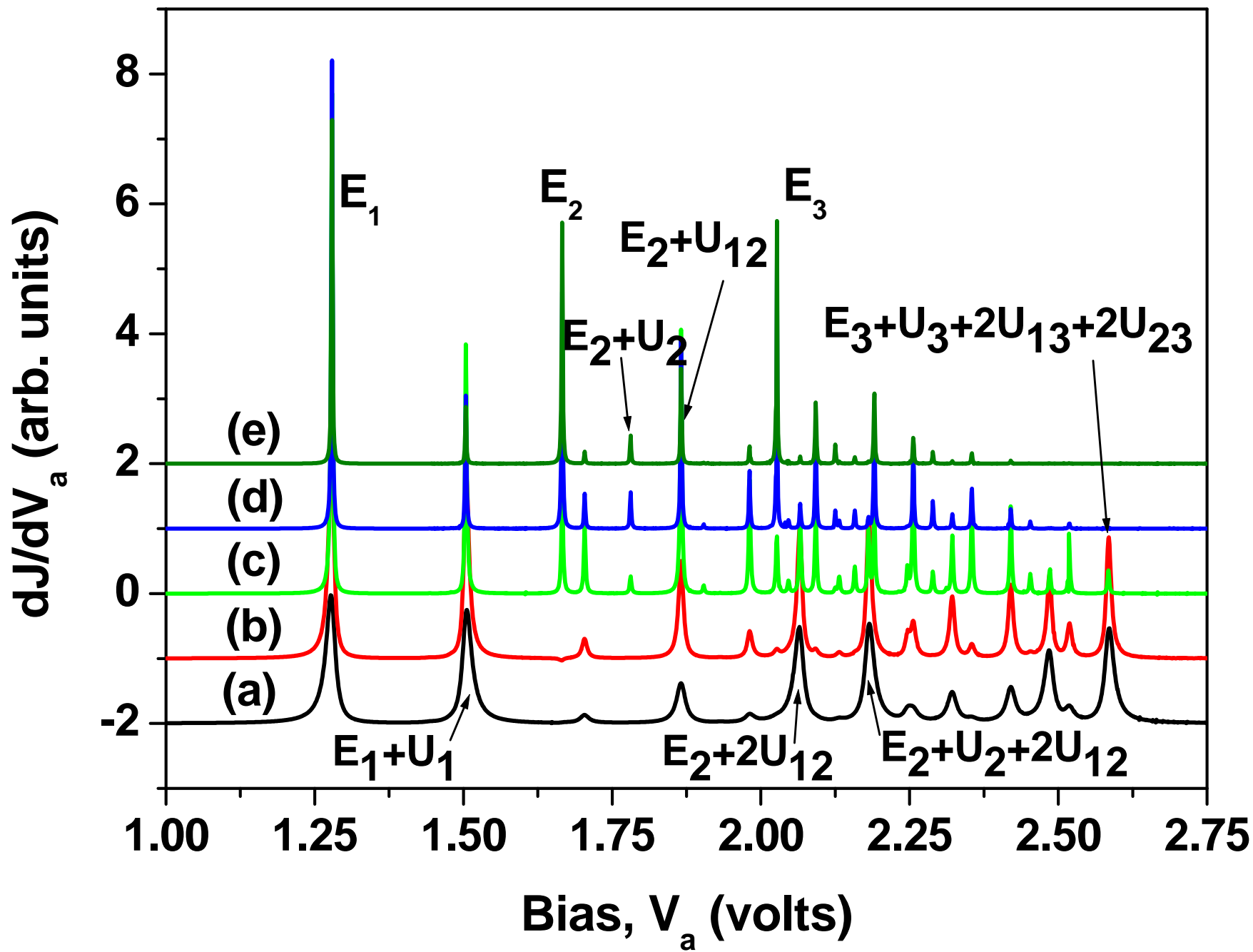
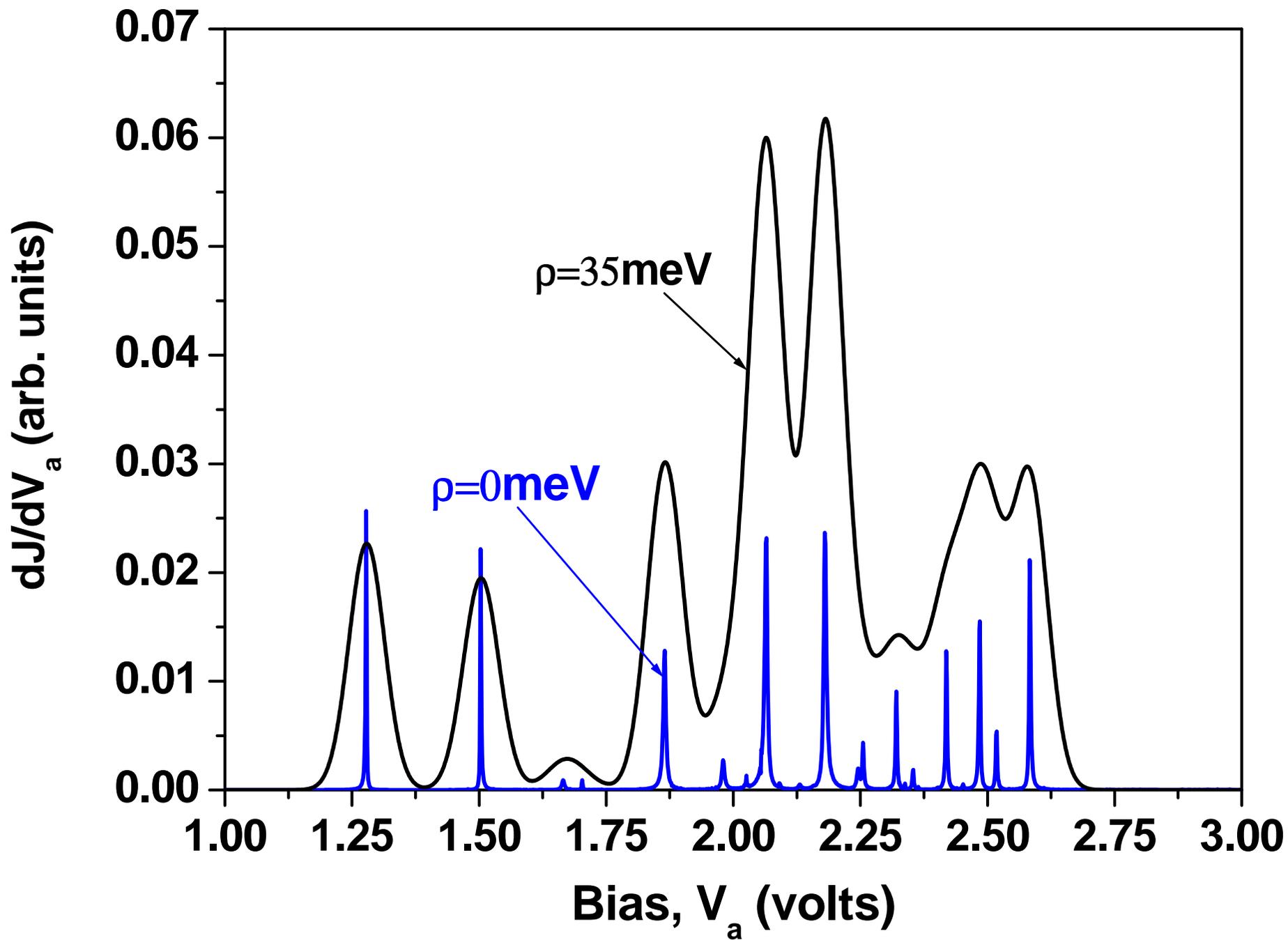
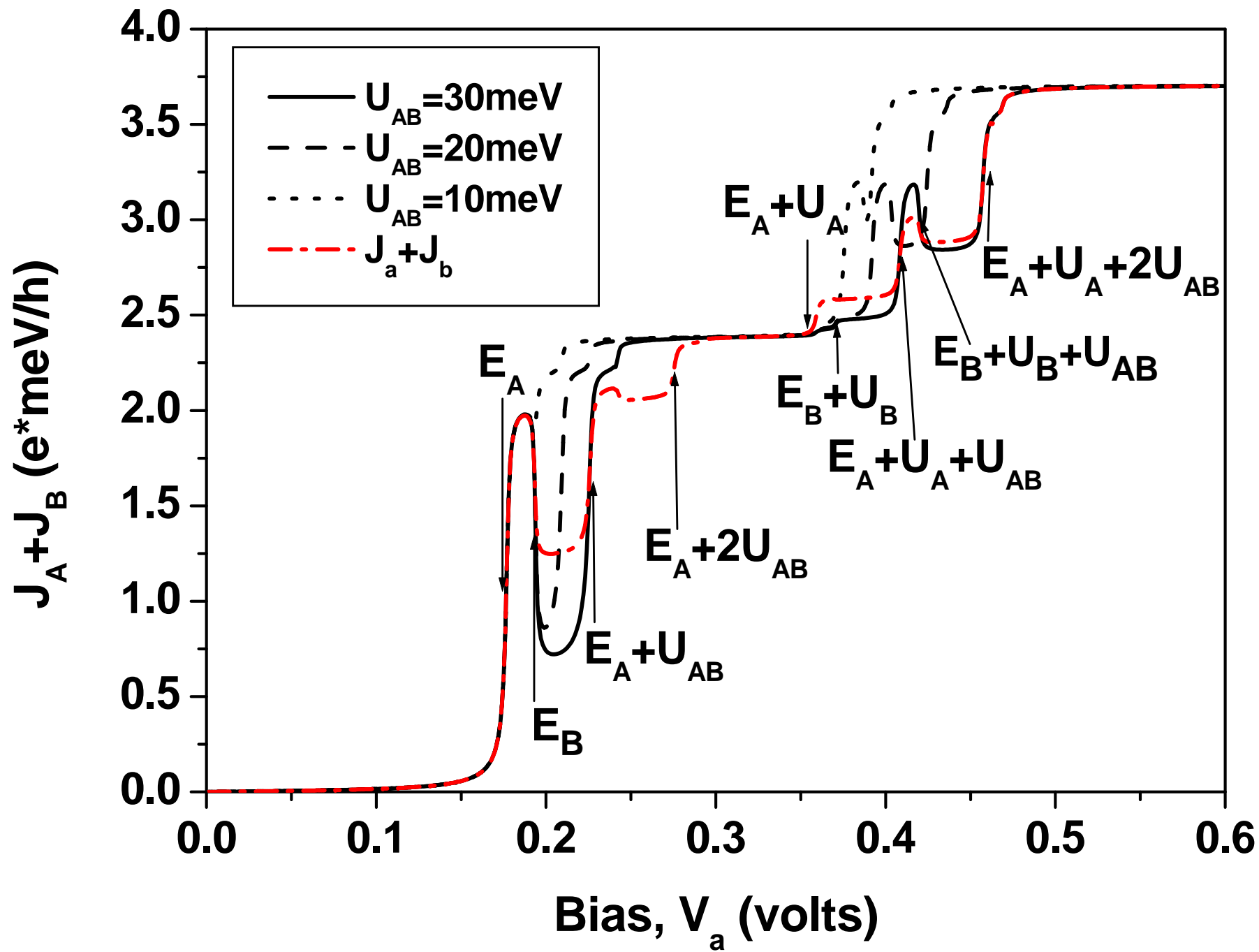
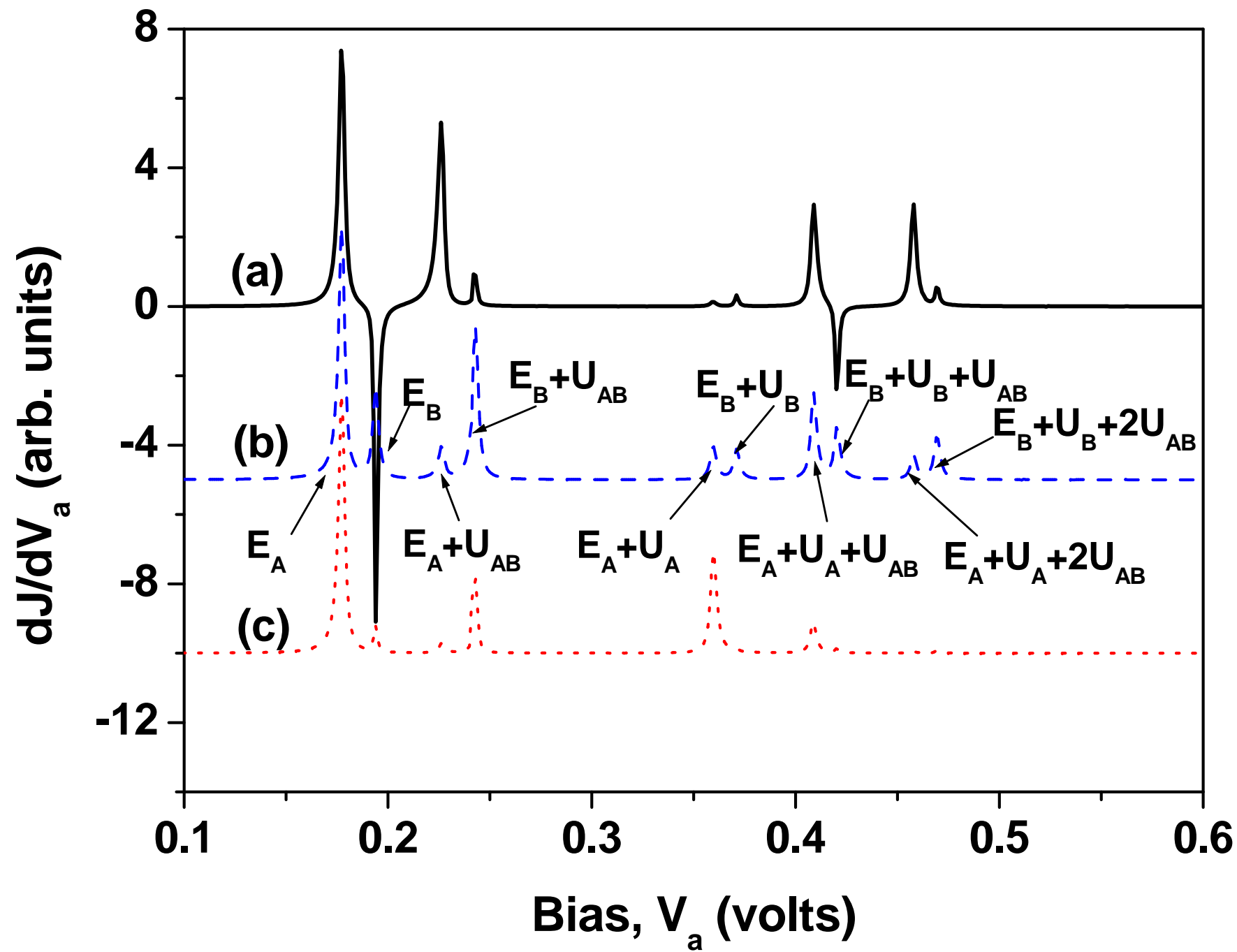
\end{document}